\documentclass[a4paper,12pt]{article}
\usepackage{amsmath}
\usepackage{amsfonts}
\usepackage{mathtools}
\usepackage{authblk}
\usepackage{graphicx,epsfig}
\usepackage{float}
\usepackage{subcaption}
\usepackage{amssymb}
\usepackage{color}

\numberwithin{equation}{section}
\title{The role of Unruh effect in Bremsstrahlung}

\author{Kajol Paithankar\footnote{kajol.paithankar@cbs.ac.in}\,}
\author{Sanved Kolekar\footnote{sanved.kolekar@cbs.ac.in} \,}
\affil{ UM-DAE Centre for Excellence in Basic Sciences,\\  
Mumbai 400098, India}

\date{March 2020}
\begin{document}

\maketitle

\begin{abstract}
An equivalence is demonstrated, by an explicit first order quantum calculation, between the Minkowski photon emission rate in the inertial frame for an accelerating charge moving on a Rindler trajectory with additional transverse drift motion and the combined Rindler photon emission and absorption rate of the same charge in the Rindler frame in the presence of the Davies Unruh thermal bath. The equivalence also extends, for the Bremsstrahlung emitted by the same charge as calculated using the machinery of classical electrodynamics. The equivalence is shown to also hold for the case of accelerating charges moving on a Rindler trajectory with additional arbitrary transverse motion. Our results generalise those of Higuchi et. al. (1992) and of Cozzella et. al. (2017) for accelerated trajectories with circular transverse motion. Related issues and experimental implications are discussed.

\end{abstract}

\section{Introduction}

Unruh effect is one of the most interesting predictions of quantum field theory when observer dependent relativistic motions are taken into account. It is well known that, an accelerated detector with constant proper acceleration $a$ in Minkowski vacuum detects a thermal bath of particles at temperature, $T_U = \hbar a/2\pi c k_B$ \cite{Unruh}. The physics of the Unruh effect in accelerated frames offers insights into those for Hawking radiation in curved spacetime.  
A direct experimental confirmation of the Unruh effect is difficult to achieve in practice since the magnitude of the constant acceleration required to observe the Unruh radiation, say at temperature $1$ K, is of the order of $10^{20}$ $m/s^2$ \cite{Review}. Numerous experimental proposals whose outcomes can be indirectly explained due to the existence of the Unruh effect have been proposed in the literature such as the electron depolarization in storage rings \cite{Bell}, decay of accelerated protons \cite{Muller}, interaction of ultraintense lasers with electrons \cite{Chen} etc. (see also \cite{Review} and references therein).

Within the framework of classical field theory, by suitably defining a classical particle number as the energy associated with a time-like Killing field divided by frequency, the number of Minkowski particles emitted by a source moving on an arbitrary trajectory was shown to be related to the number of Rindler particles emitted by the same source but with an additional thermally weighted factor to account for the presence of the Unruh bath in the Rindler frame \cite{Classical field theory}. Even though the Unruh temperature has a purely quantum origin, the quantity $\hbar \omega/ k_B T_U =  2 \pi c \omega/ a$ which appears in the weight factor is classical. In a quantum treatment up to linear order perturbations, the Minkowski photons emission rate by a uniformly accelerated point charge moving in a flat space-time as seen by the inertial observer in Minkowski vacuum state equals the emission (including absorption) rate of zero frequency Rindler photons in a thermal bath, with temperature being the Unruh temperature, when viewed from the corresponding Rindler frame of the point charge \cite{Bremsstrahlung Rindler, Bremsstrahlung Inertial}. The classical Larmor radiation of the uniformly accelerated point charge is built from zero energy Rindler modes and relates to the classical retarded solution obtained from the field expectation value in the coherent state as seen by inertial observers in the asymptotic future \cite{andre}. Also see references \cite{Zhou, Vanzella, Diaz} for related work on the relation between the classical and quantum counterparts in the context of the Unruh effect.
 
In a recent proposal by Cozzella et al. \cite{Classical ED}, this correspondence between the first order quantum emission rate in the inertial and Rindler frames (with an appropriate thermally weighted factor) as well as with the classical Larmor radiation was shown to hold for a point charge moving on a uniformly linearly accelerated trajectory but with an additional circular motion in the remaining two transverse directions. 
In particular, the rate of emission of photons with Minkowski frequencies in the inertial frame, when the charge is linearly coupled to the electromagnetic field in the Minkowski vacuum state turns out to be proportional to the emission (and absorption) rate of photons with Rindler frequencies in the co-accelerated Rindler frame, provided the charge is immersed in a thermal bath with Unruh temperature. 
The benefit of additional circular motion in the transverse directions, in this case, causes even the non-zero Rindler frequencies to contribute in the upward and downward transitions. 
A key observation made by Unruh and Wald in \cite{Unruh and Wald} vis-a-vis, the absorption of Rindler particles in the accelerated frame causing the emission of Minkowski particles in the inertial frame, is crucial to understand the correct thermal weight factor appearing in the quantum calculation in the Rindler frame.
For the same trajectory, a classical electrodynamics calculation of the spectrum of Larmor radiation by the charge in the transverse directions also matches with that of the first order perturbative quantum treatment, provided one adopts a prescribed regularization procedure. Using these correspondences, Cozzella et al. \cite{Classical ED} claim that the existence of the classical radiation then indirectly proves, through virtual confirmation, the existence of Unruh radiation and further propose an experimental set-up to detect this classical radiation. The challenges involved to perform the experiment are discussed in \cite{Quest}.

One could then ask how robust is this correspondence for trajectories other than the special trajectories assumed in the previous work mentioned above. 
Does the claim of virtual confirmation of the Unruh effect hold for a more general motion of the point charge thus relaxing some of the more taxing requirements of the experimental set-up. We address these issues in the present work. 

In section \ref{Special case}, we first investigate the case of point charge moving with uniform linear acceleration in the $t-z$ plane, where $t,z$ are the usual Minkowski co-ordinates, and additionally with a drift velocity along one of the transverse co-ordinates $x$. We show, by an explicit calculation, the said correspondence under investigation holds between the emission rate in the inertial frame and in the Rindler frame (in the presence of the Davies-Unruh bath) as well as the spectrum of Larmor radiation expected from classical electrodynamics. In section \ref{General case}, we generalise our results for trajectories having Rindler motion in the $t-z$ plane and an arbitrary motion in the remaining transverse directions. The conclusions are discussed in section \ref{discussion}.

The signature adopted is $(+,-,-,-)$ and the natural units, $k_B=c=G=\hbar=1$ are used throughout the paper.

\section{Rindler with transverse drift}\label{Special case}

In this section, we analyse the quantum and classical radiation by a point charge coupled to the electromagnetic field on a trajectory having uniform linear acceleration in the $t-z$ plane, where $t,z$ are the usual Minkowski co-ordinates, and additionally with a drift velocity $v$ along one of the transverse co-ordinates $x$. 

The trajectory can also be defined in terms of the conformal Rindler co-ordinates $(\lambda, \xi, x, y)$ covering the right wedge of the Minkowski spacetime in which the Rindler metric takes the form 
\begin{eqnarray}
ds^2 &=& e^{2 a \xi}(d\lambda^2-d\xi^2)-dx^2-dy^2, \label{Rindler metric}
\end{eqnarray}
where $\lambda$ is the usual Rindler time co-ordinate, the acceleration four vector is along the $\xi$ direction and $a$ is the proper acceleration of the co-moving Rindler observer who is at rest at $\xi = 0 = x= y$.
The Rindler coordinates $(\lambda, \xi)$ are related to Minkowski coordinates $(t, z)$ by,
\begin{equation}
t=a^{-1} e^{a\xi} \sinh(a\lambda), \;\; z=a^{-1} e^{a\xi} \cosh(a\lambda), \label{Transformation}
\end{equation}
Note the transverse co-ordinates $x$ and $y$ are same for both Minkowski and Rindler frame.
For a point charge $q$ drifting in the $x-$direction with a constant velocity $v$ in the Rindler frame, its worldline is described as $\xi=y=0$ and $x=v\lambda$. The four velocity of the charge $q$ in terms of the Rindler co-ordinates is then
\begin{eqnarray}
u^\mu_R &=& \gamma\,\left(1, 0, v, 0\right) \label{Trajectory in Rindler frame}
\end{eqnarray}
with the normalization constant being $\gamma = 1/\sqrt{1-v^2}$. Then using the transformations given by Eq.(\ref{Transformation}), the four velocity can be written in terms of the Minkowski co-ordinates $(t,z,x,y)$ as 
\begin{eqnarray}
u^\mu_M &=& \gamma \left( \cosh(a \lambda), \sinh(a \lambda), v, 0 \right) \label{Trajectory in Minkowski frame}
\end{eqnarray}
where $\lambda = \gamma \tau$ along the trajectory and  $\gamma$ is still the same normalization constant as defined in Eq.(\ref{Trajectory in Rindler frame}). It is a function of only the drift velocity parameter $v$ in the $x$ direction through $\gamma = 1/\sqrt{1-v^2}$.
The instantaneous velocity $v_z$ along the $z$ direction is $v_z = u^z_M/u^t_M =  \tanh{\gamma a \tau}$ with the range $-1 < v_z < +1$ for $-\infty < \tau < \infty$. The instantaneous velocity $v_z$ of the charge in the $z$ direction is relativistic in the large $|\tau|$ regime for the motion to be restricted in the (right) Rindler wedge and one could further choose the drift velocity $v$ in the $x$ direction to lie anywhere in the range $-1 < v < +1$. The results derived in this section \ref{Special case} and in section \ref{General case} are valid for the complete range of the velocities in the $z$ and transverse direction.

The acceleration four vector for the trajectory in Eq.(\ref{Trajectory in Minkowski frame}) turns out to be,
\begin{eqnarray}
a^\mu_M &=& \gamma^2 \left( a \sinh(a \lambda), a \cosh(a \lambda), 0, 0 \right). \label{accMinspecial}
\end{eqnarray}
with the corresponding magnitude of four acceleration vector being equal to $|a|=\sqrt{-g_{\mu \nu} a^\mu_M a^\nu_M}= a \gamma^2$. One can note as a consistency check, the expressions in Eq.(\ref{Trajectory in Minkowski frame}) and Eq.(\ref{accMinspecial}) reduce to those in the usual Rindler case when $v$ is set to zero.

Having defined the trajectory of interest, we proceed to evaluate the quantum emission rates of photons in Rindler and Minkowski frames and also the classical radiation emitted by  the accelerating charge using classical electrodynamics.

\subsection{Quantum Calculation in Rindler Frame:}\label{Special case in Rindler frame}
We consider a point charge $q$ to move on the classical Rindler trajectory with an additional transverse drift as described by Eq.(\ref{Trajectory in Rindler frame}) and coupled to the background quantised electromagnetic field in the Davies-Unruh bath. The conserved four current vector for the point charge in the Rindler frame is defined as 
\begin{eqnarray}
j^\mu_R &=& q \,\frac{u^\mu_R}{u^0_R} \,\delta(\xi) \,\delta(x-v \lambda) \,\delta(y) \label{current in Rindler for drift trajectory}
\end{eqnarray}
with the four velocity $u^\mu_R$ as in Eq.(\ref{Trajectory in Rindler frame}). The Lagrangian density for the background electromagnetic field is given by 
\begin{eqnarray}
\cal L &=& -\sqrt{-g} \left[\frac{1}{4} \,F^{\mu\nu} F_{\mu\nu} +\frac{1}{2\alpha} (\nabla^\mu A_\mu)^2\right]
\end{eqnarray}
which leads to the field equation $\nabla_\mu \nabla^\mu A_\nu=0$, working in Feynman gauge $\alpha=1$. Out of the four independent mode solutions of the field equation, $A_\mu^{(\ell,\omega,\mathbf{k_{\perp}})}$, two are pure gauge modes labelled by $\ell = 0, 3$ while the remaining two physical modes are not pure gauge and labelled by $\ell =1, 2$ satisfy the Lorenz gauge condition, $\nabla_\mu A^\mu =0$ \cite{Review,Bremsstrahlung Rindler}. The two physical modes for the Rindler metric in Eq.(\ref{Rindler metric}) can be expressed as 
\begin{eqnarray}
A^{(1,\omega,\mathbf{k_{\perp}})}_\mu (x^\nu) &=& \frac{1}{2\pi^2 k_{\perp}} \left(\frac{\sinh(\pi \omega/a)}{a}\right)^{1/2} \left(0,0,k_y f^{(\omega,\mathbf{k_{\perp}})}, -k_x f^{(\omega,\mathbf{k_{\perp}})} \right)\label{mode solution for polarization 1}\\
A^{(2,\omega,\mathbf{k_{\perp}})}_\mu (x^\nu) &=& \frac{1}{2\pi^2 k_{\perp}} \left(\frac{\sinh(\pi \omega/a)}{a}\right)^{1/2} \left(\partial_\xi f^{(\omega,\mathbf{k_{\perp}})}, -i \omega f^{(\omega,\mathbf{k_{\perp}})},0,0 \right)\label{mode solution for polarization 2}
\end{eqnarray}
where, 
$\omega$ is the frequency of Rindler photon, $\mathbf{k_{\perp}}$ is transverse momentum vector with magnitude $k_\perp =({k_x}^2+{k_y}^2)^{1/2}$ and the function $f^{(\omega,\mathbf{k_{\perp}})}$ defined as
\begin{eqnarray}
f^{(\omega,\mathbf{k_{\perp}})} &=& K_{i \omega/a}\left(\frac{k_{\perp}}{a} e^{a\xi}\right) \exp\left[\,i\,(k_x x+k_y y-\omega\lambda)\,\right]
\end{eqnarray}
where $K_\nu (z)$ is the modified Bessel function \cite{Tables of Integrals}. The quantised electromagnetic field operator can then be expressed as 
\begin{eqnarray}
\hat{A}_\mu(x^\nu) &=& \int_{-\infty}^{\infty} dk_x \int_{-\infty}^{\infty} dk_y \int_{0}^{\infty} d\omega \sum_{\ell =0}^{3}\left({\mathbf{a}}^{(i)} A_\mu^{(i)} (x^\nu) + h.c. \right)
\end{eqnarray}
where the label $(i)$ represents $(i)\equiv(\ell,\omega,\mathbf{k_{\perp}})$ and ($\mathbf{a}^{\dagger}_{(i)}, \mathbf{a}^{}_{(i)}$) are the corresponding creation and annihilation operators respectively satisfying the commutation relations,
\begin{eqnarray}
\left[ \mathbf{a}^{}_{(\ell,\omega,\mathbf{k_{\perp}})},\mathbf{a}^{\dagger}_{(\ell',\omega',\mathbf{k'_{\perp}})} \right] &=& \delta_{\ell \ell'}\, \delta(\omega-\omega') \,\delta(\mathbf{k_{\perp}}-\mathbf{k'_{\perp}})
\end{eqnarray}
for $\ell$ and $\ell'$ corresponding to only the physical modes. The interaction between the charged particle and the electromagnetic field can now be described by the Lagrangian density,
\begin{eqnarray}
{\cal L}_{int} &=& \sqrt{-g} \, j^\mu_R \, \hat{A}_\mu.
\end{eqnarray}
Now, to lowest order in perturbation, the amplitude for absorption of a Rindler photon to the Rindler vacuum where the photon is described by the single photon state $|\ell,\omega,\mathbf{k_{\perp}}\rangle_R= \mathbf{a}^{\dagger}_{(\ell,\omega,\mathbf{k_{\perp}})} |0\rangle_R$ is
\begin{eqnarray}
{\cal A}^{abs}_{(\ell,\omega,\mathbf{k_{\perp}})} &=& i \int d^4 x \,\sqrt{-g} \, j^\mu_R \, \prescript{}{R}{ \langle 0|} \hat{A}_\mu |\ell,\omega,\mathbf{k_{\perp}}{\rangle}_R \label{Amplitude of absorption}
\end{eqnarray}
Here, the Rindler vacuum $|0\rangle_R$ is taken to be the state annihilated by all annihilation operators $\mathbf{a}^{}_{(i)}$, that is defined as $\mathbf{a}^{}_{(\ell,\omega,\mathbf{k_{\perp}})} |0\rangle_R=0$.
Then for the four current vector $j^\mu_R$ in Eq.(\ref{current in Rindler for drift trajectory}) for the case of Rindler trajectory with a transverse drift, the absorption amplitudes for the physical modes $\ell=1,2$ are obtained to be,
\begin{eqnarray}
{\cal A}^{abs}_{(1,\omega,\mathbf{k_{\perp}})} &=& i \int d^4 x \,\sqrt{-g} \, j^x_R \, A_x^{(1,\omega,\mathbf{k_{\perp}})}\\
 &=& \frac{i q}{\pi} \left(\frac{\sinh(\pi \omega/a)}{a}\right)^{1/2} \left(\frac{v k_y}{k_\perp}\right) K_{i \omega/a}\left(\frac{k_{\perp}}{a}\right) \delta(\omega-k_x v)\label{Amplitude with x component}\\
{\cal A}^{abs}_{(2,\omega,\mathbf{k_{\perp}})} &=& i \int d^4 x \,\sqrt{-g} \, j^\lambda_R \, A_{\lambda}^{(2,\omega,\mathbf{k_{\perp}})}\\
 &=& \frac{i q}{\pi} \left(\frac{\sinh(\pi \omega/a)}{a}\right)^{1/2} K'_{i \omega/a}\left(\frac{k_{\perp}}{a}\right) \delta(\omega-k_x v)\label{Amplitude with lambda component}
\end{eqnarray}
where, prime denotes the derivative with respect to the argument of Bessel function. 

We shall assume that the state of the quantised electromagnetic field is the Minkowski vacuum state or equivalently the Davies-Unruh thermal bath in the Rindler wedge with temperature $T = a/2\pi$. In such a case, the probability of absorption is additionally weighed by the thermal factor $1/[\exp{(\omega/T)} - 1]$ corresponding to the number of photons already present in the background thermal bath, for each Rindler photon frequency $\omega$. Thus the total rate of absorption of Rindler photons is then 
\begin{eqnarray}
{P}^{abs}_{R} &=& \sum_{\ell =1,2}\int_{-\infty}^{\infty} d k_x \int_{-\infty}^{\infty} d k_y \int_0^\infty d\omega \, \frac{|{\cal A}^{abs}_{(\ell,\omega,\mathbf{k_{\perp}})}|^2}{\Delta \tau} \, \left( \frac{1}{e^{\omega/T} - 1} \right)
\label{absrate}
\end{eqnarray}
where $\Delta \tau$ is the total proper time interval of Rindler observer during which the interaction remains switched on. Since we also have, $|{\cal A}^{abs}_{(\ell,\omega,\mathbf{k_{\perp}})}|=|{\cal A}^{emi}_{(\ell,\omega,\mathbf{k_{\perp}})}|$, the total emission rate of Rindler photons is found by a similar procedure to be
\begin{eqnarray}
{P}^{emi}_{R} &=& \sum_{\ell =1,2}\int_{-\infty}^{\infty} d k_x \int_{-\infty}^{\infty} d k_y \int_0^\infty d\omega \, \frac{|{\cal A}^{abs}_{(\ell,\omega,\mathbf{k_{\perp}})}|^2}{\Delta \tau} \, \left( 1 + \frac{1}{e^{\omega/T} - 1} \right)
\label{emirate}
\end{eqnarray}
where the factor of unity in the last term in the bracket corresponds to spontaneous emission. Thus the total rate which includes the emission rate as well as the absorption rate is given by,
\begin{eqnarray}
{P}^{total}_{R} &=& \sum_{\ell =1,2}\int_{-\infty}^{\infty} d k_x \int_{-\infty}^{\infty} d k_y \int_0^\infty d\omega \, \frac{|{\cal A}^{abs}_{(\ell,\omega,\mathbf{k_{\perp}})}|^2}{\Delta \tau} \, \coth\left(\frac{\omega}{2T}\right)
\label{totalrate}
\end{eqnarray}
The emission and absorption rates in Eq.(\ref{emirate}) and (\ref{absrate}) are added to arrive at the $\coth(\omega /2T )$ factor in the total rate ${P}^{total}_{R} $. Such a reasoning, as elaborated in \cite{Bremsstrahlung Rindler, Classical ED}, is based on Unruh and Wald's observation in \cite{Unruh and Wald} that the absorption of Rindler particles in the presence of the background Davies-Unruh bath in the accelerated frame is seen by the inertial observer as the emission of Minkowski particles in the inertial frame.

In the expression of ${P}^{total}_{R} $ in Eq.(\ref{totalrate}), the amplitude of absorption is proportional to $\delta(\omega-k_x v)$ and one can simply evaluate the integral over $\omega$. The Rindler photon energy $\omega$ can only have non-negative values in range $[0,\infty)$, which restricts $k_x$ to be non negative since we can choose $v$ to be positive (a particular direction of drift which in this case is the positive $x$ direction). Thus, the total rate is obtained as,
\begin{eqnarray}
{P}^{total}_{R} &=& \frac{q^2}{{2\pi}^3 a} \int_{-\infty}^{\infty} d k_x \int_{-\infty}^{\infty} d k_y \,\Theta(k_x) \sinh\left(\frac{\pi k_x v}{a}\right) \coth\left(\frac{k_x v}{2 T}\right)\nonumber\\
& & \,\times\left(\left|K'_{ik_x v/a}\left(k_{\perp}/a\right)\right|^2+\left(\frac{v k_y}{k_\perp}\right)^2\left|K_{ik_x v/a}\left(k_{\perp}/a\right)\right|^2\right)
\end{eqnarray}
where, $\Theta(k_x)$ is the Heavyside step function 
and we have identified $\Delta \tau= 2 \pi \delta(0)$ as per the regularisation procedure adopted in \cite{Bremsstrahlung Rindler, Classical ED}. In the above expression, we now set the temperature T of the background bath to be equal to the Unruh bath temperature $T_U = a/2\pi$ to finally get 
\begin{eqnarray}
{P}^{total}_{R} &=& \frac{q^2}{{2\pi}^3 a} \int_{0}^{\infty} d k_x \int_{-\infty}^{\infty} d k_y \,\cosh\left(\frac{\pi k_x v}{a}\right) \times\nonumber\\
& & \left(\left|K'_{ik_x v/a}\left(k_{\perp}/a\right)\right|^2+\left(\frac{v k_y}{k_\perp}\right)^2\left|K_{ik_x v/a}\left(k_{\perp}/a\right)\right|^2\right)\label{Rindler Transition rate}
\end{eqnarray}
Substituting $v=0$, one can check the consistency of the above expression with the total rate obtained in the case of the zero frequency modes in \cite{Bremsstrahlung Rindler}.

\subsection{Quantum Calculation in Minkowski Frame:}\label{Special case in Minkowski frame}
In this subsection, we calculate the emission rate in the inertial frame for the same point charge $q$ as described by the trajectory in Eq.(\ref{Trajectory in Rindler frame}) and coupled to the background quantised electromagnetic field in the Minkowski vacuum state. In terms of the Minkowski co-ordinates, the corresponding conserved four vector current is
\begin{eqnarray}
j^\mu_M &=& q \,\frac{u^\mu_M}{u^0_M}\, \delta\left(z-a^{-1}\cosh(a\lambda)\right) \delta(x-v \lambda) \delta(y).\label{current in inertial frame}
\end{eqnarray}
The quantised electromagnetic field in the inertial frame is expressed in terms of the standard Minkowski plane wave mode solutions and given by \cite{Bremsstrahlung Inertial},
\begin{eqnarray}
\hat{A}_\mu(x) &=& \int \frac{d^3 \mathbf{k}}{2 k_0 (2\pi)^3} \sum_{\ell=1}^4 \left[a^{(\ell)}\epsilon_\mu^{(\ell)} e^{-i k_\nu x^\nu}+H.c.\right]
\end{eqnarray}
where $k_0=\sqrt{k_z^2+k_\perp^2}$ is the energy of Minkowski photon and $\epsilon_\mu^{(\ell)}$ are polarization vectors. As in the Rindler frame case, we label the two pure gauge modes out of the four independent mode solutions by $\ell = 0, 3$, and the two physical modes by $\ell =1, 2$. The two physical modes satisfy the Lorenz gauge condition, $\nabla_\mu A^\mu =0$. Accordingly the polarisation vectors $\epsilon_\mu^{(\ell)}$ are chosen in a Cartesian frame such that $k^\mu =\left(|k|,0,0,|k|\right)$ as in \cite{Bremsstrahlung Inertial}. 

The amplitude of emission of a single Minkowski photon with momentum ${\mathbf{k}}$ and polarization $\ell$ to the background Minkowski vacuum state can be computed as 
\begin{eqnarray}
{\cal A}^{em}_{(\ell,{\mathbf{k}})} &=& i \int d^4 x \,j^\mu_M (x) \, \prescript{}{M}\langle {\mathbf{k}},\ell | \, \hat{A}_\mu (x)\, |0\rangle_{M} 
\end{eqnarray}
Hence the rate of emission of photons with transverse momentum $\mathbf{k_\perp}$ then becomes,
\begin{eqnarray}
P^{em}_{M,\mathbf{k_\perp}} &=& \frac{1}{\Delta\tau} \sum_{\ell =1}^{2} \int_{-\infty}^{\infty}\frac{d k_z}{2 k_0 (2\pi)^3} |{\cal A}^{em}_{(\ell,{\mathbf{k}})}|^2\\
 &=& \frac{-1}{\Delta\tau} \int_{-\infty}^{\infty}\frac{dk_z}{2 k_0 (2\pi)^3} \int d^4 x \int d^4 x' j^\mu(x)\, j_\mu(x') e^{i k_0(t-t')-i\mathbf{k}\cdot(\mathbf{x}-\mathbf{x'})}\label{General expression for emission rate}
\end{eqnarray}
where, we have dropped the subscript $M$ of the current $j^\mu_M$ in the above expression for calculational simplicity.  Substituting the current $j^\mu$ from Eq.(\ref{current in inertial frame}) and performing the spatial integrals, we get,
\begin{eqnarray}
P^{em}_{M,\mathbf{k_\perp}} &=& \frac{-q^2}{2\Delta\tau} \int_{-\infty}^{\infty}\frac{d k_z}{2k_0 (2\pi)^3} \int_{-\infty}^{\infty} d\sigma \int_{-\infty}^{\infty} d\rho \left(\cosh(a\sigma)-v^2\right) e^{-i k_x v \sigma}\nonumber\\
 & & \times \,\exp\left[\frac{2i}{a} \sinh\left(\frac{a\sigma}{2}\right)\left(k_0\cosh\left(\frac{a\rho}{2}\right)-k_z \sinh\left(\frac{a\rho}{2}\right)\right)\right] 
\end{eqnarray}
where, $\sigma=\lambda-\lambda'$ and $\rho=\lambda+\lambda'$. To evaluate the above integral, we proceed in a similar way to \cite{Bremsstrahlung Inertial}. We define new variables $\bar{k_0}$ and $\bar{k_z}$  by using the following transformations,
\begin{eqnarray}
\bar{k_0} &=& k_0 \cosh(a\rho/2)- k_z \sinh(a\rho/2)\label{transformed frequency}\\
\bar{k_z} &=& k_z \cosh(a\rho/2)- k_0 \sinh(a\rho/2)\label{transformed momentum}
\end{eqnarray}
with $\bar{k_0}=\sqrt{\bar{k_z}^2+k_\perp^2}$, which essentially boosts back the momentum variables. Doing so, essentially makes the integral independent of $\rho$ and we can factor out infinite integral $\int_{-\infty}^{\infty} d(\lambda+\lambda')/2=\int_{-\infty}^{\infty} d\rho/2$ by identifying $\int_{-\infty}^{\infty} d\rho/2 = \Delta\tau$, the total proper time interval as the Minkowski regularisation adopted in \cite{Bremsstrahlung Inertial}. The resulting simplified expression is 
\begin{eqnarray}
P^{em}_{M,\mathbf{k_\perp}} &=& -q^2 \int_{-\infty}^{\infty}\frac{d \bar{k_z}}{2\bar{k_0} (2\pi)^3} \int_{-\infty}^{\infty} d\sigma \left(\cosh(a\sigma)-v^2\right) e^{-i k_x v \sigma}\nonumber\\
 & & \times \,\exp\left[\frac{2i\bar{k_0}}{a} \sinh\left(\frac{a\sigma}{2}\right)\right]
\end{eqnarray}
We next define new variables $S_+$ and $S_-$ as per the following relations
\begin{eqnarray}
S_\pm &=& \frac{\bar{k_0}+\bar{k_z}}{k_\perp}\; e^{\pm a\sigma/2}
\end{eqnarray}
In terms of these variables, one can express the integral as
\begin{eqnarray}
P^{em}_{M,\mathbf{k_\perp}} &=& \frac{-q^2}{4a (2\pi)^3} \int_{0}^{\infty} dS_+ \int_{0}^{\infty} dS_- \left(\frac{1}{S_+^2}+\frac{1}{S_-^2}-\frac{2v^2}{S_+S_-}\right) \left(\frac{S_+}{S_-}\right)^{-i k_x v/a}\nonumber\\
 & & \times \exp\left[\frac{i k_\perp}{2a}\left(S_+-\frac{1}{S_+}\right)\right] \exp\left[\frac{-i k_\perp}{2a}\left(S_--\frac{1}{S_-}\right)\right]
\end{eqnarray}
Using the integral representation of modified Bessel function \cite{Tables of Integrals} and and their recurrence relations,
\begin{eqnarray}
\int_{0}^{\infty} t^{\nu-1} \exp\left[\frac{i x}{2}\left(t-\frac{z^2}{t}\right)\right] dt &=& 2\, z^\nu \,e^{i\pi\nu/2}\, K_\nu(xz) \label{Integral representation of Bessel function}
\end{eqnarray}
the integrals are evaluated to get the emission rate for fixed transverse momenta $k_x$ and $k_y$ to be 
\begin{eqnarray}
P^{em}_{M,\mathbf{k_\perp}} &=& \frac{q^2}{4 \pi^3 a} \exp\left[\frac{\pi k_x v}{a}\right] \left[\left|K'_{ik_x v/a}\left(k_{\perp}/a\right)\right|^2+\left(\frac{v k_y}{k_\perp}\right)^2\left|K_{ik_x v/a}\left(k_{\perp}/a\right)\right|^2\right]
\label{emiratemin}
\end{eqnarray}
Hence total emission rate is written by integrating over all the transverse modes to get
\begin{eqnarray}
P^{em}_{M} &=& \frac{q^2}{2 \pi^3 a} \int_{0}^{\infty} dk_x \int_{-\infty}^{\infty} dk_y \,\cosh\left(\frac{\pi k_x v}{a}\right) \times \nonumber\\
 & & \left(\left|K'_{ik_x v/a}\left(k_{\perp}/a\right)\right|^2+\left(\frac{v k_y}{k_\perp}\right)^2\left|K_{ik_x v/a}\left(k_{\perp}/a\right)\right|^2\right)
 \label{Inertial transition rate}
\end{eqnarray}
which is equivalent to the total rate computed in the Rindler frame in the presence of the Davies-Unruh bath in Eq.(\ref{Rindler Transition rate}) as expected.

\subsection{Radiation using classical electrodynamics}\label{Classical calculation of special case}
In this subsection, we perform a simple classical electrodynamics calculation involving the spectral angular distribution of radiation from an accelerating charge on the Rindler trajectory with transverse drift. The number of emitted photons, each of energy $k_0$, is given by the expression
\begin{eqnarray}
N_{M} &=& \int_{0}^{\infty} \frac{dk_0}{k_0} \int^{\pi}_{0} \sin{\theta}\, d\theta \int_{0}^{2\pi} d\phi \; I(k_0,\theta,\phi)\label{Total number of emitted photons}
\end{eqnarray}
where, $I(k_0,\theta,\phi)$ is the spectral angular distribution which is the energy emitted through classical radiation by the accelerating charge for a particular frequency $k_0$ in a particular direction $\theta, \phi$. Then changing the variables $(k_0,\theta,\phi)$ from spherical co-ordinates to the Cartesian ones $(k_x,k_y,k_z)$ and noting that the volume element changes as $dk_0\, d(\cos\theta)\, d\phi =k_0^{-2}\, dk_x\, dk_y\, dk_z$ with $k_0=\sqrt{k_x^2+k_y^2+k_z^2}$ , the total number of emitted photons with the fixed transverse momenta $k_x$, $k_y$ can be written as,
\begin{eqnarray}
N_{M,\mathbf{k_\perp}} &=& \int_{-\infty}^{\infty} \frac{dk_z}{k_0^3}\; I(k_x,k_y,k_z)\label{Number of emitted photons}
\end{eqnarray}

We shall demonstrate below that the photon number $N_{M,\mathbf{k_\perp}}$ is proportional to the total emission rate ${P}^{em}_{M,\mathbf{k_\perp}}$ obtained through the quantum calculation in the inertial frame as expressed in Eq.(\ref{emiratemin}). For an accelerated point charge $q$, the spectral angular distribution $I(k_0,\theta,\phi)$ using standard classical electrodynamics is given by \cite{Zangwill},
\begin{eqnarray}
I(k_0,\theta,\phi) &=& \frac{q^2k_0^2}{4\pi^2} \, {\left|\bar{F}(k_0,\theta,\phi)\right|}^2 \label{Spectal angular distribution}
\end{eqnarray}
where the function $\bar{F}(k_0,\theta,\phi)$ is defined as
\begin{eqnarray}
\bar{F}(k_0,\theta,\phi) &=& \hat{r}\times\int_{-\infty}^{\infty} d\lambda\, \frac{d\bar{r_q}}{d\lambda}\, \exp\left[\,-ik_0\,\hat{r}\cdot\bar{r_q}(\lambda)+ik_0 \,t(\lambda)\,\right]\label{Vector F}
\end{eqnarray}
where, $\hat{r}=\bar{k}/k_0$ is the unit vector pointing in the observed direction and $\bar{r_q}= v\lambda \,\hat{i}+a^{-1} \cosh(a\lambda)\, \hat{k}$, is the trajectory, in the spatial Cartesian coordinates, of the charge $q$ having Rindler motion with additional transverse drift and the Minkowski time co-ordinate $t(\lambda)=a^{-1} \sinh(a\lambda)$ is expressed in terms of other parameter $\lambda$ (which coincides with the Rindler time co-ordinate). Substituting the trajectory in the above expression, we get,
\begin{eqnarray}
\bar{F}(k_x,k_y,k_z) &=& \int_{-\infty}^{\infty} d\lambda\, \left[\frac{k_y \sinh(a\lambda)}{k_0} \,\hat{i} + \left(\frac{k_z v}{k_0}-\frac{k_x \sinh(a\lambda)}{k_0}\right)\, \hat{j} - \frac{k_y v}{k_0}\, \hat{k}\right]\nonumber\\
& & \quad \quad\times \,\exp\left[\frac{i k_0 \sinh(a\lambda)}{a}- \frac{i k_z \cosh(a\lambda)}{a}-i k_x v \lambda\right]
\end{eqnarray}
Using the integral representation of Bessel function as given in Eq.(\ref{Integral representation of Bessel function}), the integral in above expression can be simplified to arrive at the following form for ${\left|\bar{F} (k_x,k_y,k_z)\right|}^2$,
\begin{eqnarray}
{\left|\bar{F}(k_x,k_y,k_z)\right|}^2 &=& \frac{4}{a^2} \exp\left[\frac{\pi k_x v}{a}\right] \times \nonumber\\
 & & \left(\left|K'_{ik_x v/a}\left(k_{\perp}/a\right)\right|^2+\left(\frac{v k_y}{k_\perp}\right)^2\left|K_{ik_x v/a}\left(k_{\perp}/a\right)\right|^2\right)
\end{eqnarray}  
Using the above expression in Eq.(\ref{Spectal angular distribution}), the number of emitted photons $N_{M,\mathbf{k_\perp}}$ with the fixed transverse momentum $\mathbf{k_\perp}$ in Eq.(\ref{Number of emitted photons}) is found to be
\begin{eqnarray}
N_{M,\mathbf{k_\perp}} &=& \frac{q^2}{\pi^2 a^2} \exp\left(\frac{\pi k_x v}{a}\right) \left(\left|K'_{ik_x v/a}\left(k_{\perp}/a\right)\right|^2+\left(\frac{v k_y}{k_\perp}\right)^2\left|K_{ik_x v/a}\left(k_{\perp}/a\right)\right|^2\right) \nonumber\\
 & & \quad\times \int_{-\infty}^{\infty}\frac{dk_z}{{({k_\perp^2+k_z^2})}^{1/2}}
\end{eqnarray}
One can note here that the number of emitted photons with a fixed transverse momentum $N_{M,\mathbf{k_\perp}}$ is proportional to the quantum emission rate with a fixed transverse momentum $P^{em}_{M,\mathbf{k_\perp}}$ given by Eq.(\ref{emiratemin}) in the inertial frame with a divergent proportionality constant.

We can next identify the divergent integral over $k_z$ with the infinite amount of time $\Delta\tau$, the charge is assumed to be accelerating for \cite{Classical ED}. In such a case, the number of photons emitted with a fixed $\mathbf{k_\perp}$ will be infinite. The total number of emitted photons is then expressed as 
\begin{eqnarray}
N_{M,\mathbf{k_\perp}} &=&  \frac{q^2}{\pi^2 a} \Delta\tau \exp\left(\frac{\pi k_x v}{a}\right) \nonumber\\
& &\quad\times\left(\left|K'_{ik_x v/a}\left(k_{\perp}/a\right)\right|^2+\left(\frac{v k_y}{k_\perp}\right)^2\left|K_{ik_x v/a}\left(k_{\perp}/a\right)\right|^2\right)\\
 &=& 4\pi \,\Delta\tau \,{P}^{em}_{M,\mathbf{k_\perp}}
\end{eqnarray}
and is proportional to the emission rate computed in the inertial frame in Eq.(\ref{emiratemin}) with proportionality constant equal to $4\pi \,\Delta\tau$.

\section{Rindler with arbitrary transverse motion}\label{General case}

In this section, we generalise the results derived in the previous section for Rindler trajectories having an additional arbitrary transverse motion. 

Below, we describe the trajectory of the point charge $q$. We consider the charge to have the usual Rindler motion in $t-z$ plane while having an arbitrary motion in transverse directions $x$ and $y$. We again define the trajectory first in terms of the conformal co-ordinates with the Rindler metric in the right Rindler wedge, 
\begin{eqnarray}
ds^2 &=& e^{2 a \xi}(d\lambda^2-d\xi^2)-dx^2-dy^2\label{Rindler metric 2},
\end{eqnarray}
where $\lambda$ is the usual Rindler time co-ordinate, $a$ is the proper acceleration of the co-moving Rindler observer at rest at $\xi = 0 = x= y$ and whose acceleration four vector is along the $\xi$ direction.

We define the class of trajectories with Rindler motion in the $\lambda - \xi$ plane, $\xi(\lambda) = 0$ and arbitrary motion in the transverse directions with $x(\lambda)$ and $y(\lambda)$ by the four velocity vector,
\begin{eqnarray}
\tilde{u}^\mu_R &=& \gamma\,\left(\,1, 0, u_x(\lambda),u_y(\lambda)\,\right)=\gamma \,u^\mu_R\label{velocity in Rindler frame for general trajectory}
\end{eqnarray}
where the normalisation factor $\gamma = 1/\sqrt{1-u_x^2-u_y^2}$ is, in general, not a constant. Thus the proper time $\tau$ along the trajectory considered may not be proportional to the Rindler time co-ordinate $\lambda$ as in the earlier case and satisfies $d\lambda/ d\tau = \gamma$. In terms of the Minkowski co-ordinates in the inertial frame, the corresponding four velocity reads as,
\begin{eqnarray}
\tilde{u}^\mu_M &=& \gamma \left(\,\cosh(a \lambda),\sinh(a \lambda),u_x(\lambda),u_y(\lambda)\,\right) = \gamma\, u^\mu_M\label{velocity in Minkowski frame for general trajectory}
\end{eqnarray}
The magnitude of proper acceleration for this trajectory is, $|a|^2= \gamma^4 (a^2+a_x^2+a_y^2)-3\gamma^6 (u_x a_x+u_y a_y)^2$, where $a_{(x,y)} =du_{(x,y)}/d\lambda$.
Thus the proper acceleration for these trajectories is in general proper time dependent. 

The transverse components of the four velocity are, in general, arbitrary smooth functions of $\lambda$. In the special case, when $u_x = \sinh(a \lambda)$, the charge will also undergo a corresponding Rindler motion in the $t-x$ plane. Then, due to the explicit symmetry in the motion of the charge in the $z$ and $x$ directions, one could also perform an explicit calculation of the quantum rates in the rest frame of the Rindler motion in the $t-x$ plane (instead of the rest frame of the Rindler motion in the $t-z$ plane as chosen in Eq.(\ref{Rindler metric 2})), with $z$ now being one of the transverse directions. The roles of $z$ and $x$ would be simply exchanged. In such a case, one would again have an Unruh effect due to the Rindler motion in the $t-x$ plane which allows to define the background thermal state for a quantum calculation. The conclusions regarding the equivalence in the corresponding rates in this case will be exactly same as those obtained in the sections below.

With the trajectory defined, we now proceed to obtain the expressions for the photon emission rates in the inertial frame and Rindler frame with a thermal bath and also the bremsstrahlung using classical electrodynamics.

\subsection{Quantum calculation in the Minkowski frame}\label{General case in Minkowski frame}
In this subsection, we calculate the emission rate in the inertial frame for a point charge $q$ as described by the trajectory in Eq.(\ref{velocity in Minkowski frame for general trajectory}) and coupled to the background quantised electromagnetic field in the Minkowski vacuum state. In terms of the Minkowski co-ordinates, the corresponding conserved four vector current is
\begin{eqnarray}
j^\mu_M &=& q \,\frac{\tilde{u}^\mu_M}{\tilde{u}^0_M} \,\delta(z-a^{-1} \cosh(a\lambda)) \,\delta\left(x-x(\lambda)\right) \,\delta\left(y-y(\lambda)\right)\\
 &=& q \,\frac{u^\mu_M}{u^0_M} \,\delta(z-a^{-1} \cosh(a\lambda)) \,\delta\left(x-x(\lambda)\right) \,\delta\left(y-y(\lambda)\right)\label{current in Minkowski for general trajectory}
\end{eqnarray}
The rate of emission of photons with a fixed transverse momentum $\mathbf{k_\perp}$ as measured in an inertial frame is then given by Eq.(\ref{General expression for emission rate}). Substituting the four current $j^\mu$ from Eq.(\ref{current in Minkowski for general trajectory}) in Eq.(\ref{General expression for emission rate}) and performing the spatial integrals, we get,
\begin{eqnarray}
P^{total}_{M,\mathbf{k_\perp}} &=& \frac{-q^2}{\Delta \tau} \int_{-\infty}^{\infty}\frac{d \bar{k_z}}{2\bar{k_0} (2\pi)^3} \int_{-\infty}^{\infty} d\lambda \int_{-\infty}^{\infty} d\lambda' \left[u^\mu_M(\lambda){u_M}_\mu(\lambda')\right] \nonumber\\
 & & \times \,e^{-i \mathbf{k_{\perp}}\cdot(\mathbf{x_{\perp}}-\mathbf{x'_{\perp}})} \exp\left(\frac{2i\bar{k_0}}{a} \sinh\left(\frac{a\sigma}{2}\right)\right)
\end{eqnarray}
where, $\mathbf{x_{\perp}} \equiv (\,x(\lambda),y(\lambda)\,)$ and $\bar{k_0}$ is as defined in Eq.(\ref{transformed frequency}) with $\sigma=\lambda-\lambda'$ and $\rho=\lambda+\lambda'$.
The overall emission rate, then can be obtained to be,
\begin{eqnarray}
P^{total}_{M}  &=& \int_{-\infty}^{\infty} dk_x \int_{-\infty}^{\infty} d k_y \,P^{total}_{M,\mathbf{k_\perp}}\\
 &=& \frac{-q^2}{\Delta \tau} \int_{-\infty}^{\infty} dk_x \int_{-\infty}^{\infty} d k_y \int_{-\infty}^{\infty}\frac{d \bar{k_z}}{2\bar{k_0} (2\pi)^3} \int_{-\infty}^{\infty} d\lambda  \nonumber\\
& &  \times \int_{-\infty}^{\infty} d\lambda'\left[ \left[u^\mu(\lambda)u_\mu(\lambda')\right]\, e^{-i \mathbf{k_{\perp}}\cdot(\mathbf{x_{\perp}}-\mathbf{x'_{\perp}})}\, \exp\left(\frac{2i\bar{k_0}}{a} \sinh\left(\frac{a\sigma}{2}\right)\right)\right]\label{Emission rate for general trajectory}
\end{eqnarray}
Here we have dropped the label $M$ for the four velocity $u^\mu_M$ for notational simplicity. From here onwards $u^\mu$ shall represent the Minkowski four velocity as specified in Eq.(\ref{velocity in Minkowski frame for general trajectory}) unless otherwise stated. Now, we proceed to get the expression for the total rate for a Rindler detector.

\subsection{Quantum calculation in the Rindler frame}\label{General case in Rindler frame}
In this subsection, we calculate the emission rate in the Rindler frame for a point charge $q$ moving on the Rindler trajectory with additional arbitrary transverse motion as described by Eq.(\ref{velocity in Rindler frame for general trajectory}) and coupled to the background quantised electromagnetic field in the Davies-Unruh bath. The conserved four current vector $j^\mu_R$ for the point charge in the Rindler frame is defined as 
\begin{eqnarray}
j^\mu_R &=& q \,\frac{\tilde{u}^\mu_R}{\tilde{u}^0_R} \,\delta(\xi) \,\delta\left(x-x(\lambda)\right) \,\delta\left(y-y(\lambda)\right)\\
 &=& q \,\frac{u^\mu_R}{u^0_R} \,\delta(\xi) \,\delta\left(x-x(\lambda)\right) \,\delta\left(y-y(\lambda)\right)
\label{current in Rindler for general trajectory}
\end{eqnarray}
For the above four current, the amplitude of absorption of a single Rindler photon as defined by Eq.(\ref{Amplitude of absorption}) can be computed by using the mode solutions in Rindler frame, Eq.(\ref{mode solution for polarization 1}) and Eq.(\ref{mode solution for polarization 2}). The amplitudes of absorption corresponding to the two physical modes $\ell=1,2$, then turn out to be,
\begin{eqnarray}
{\cal A}^{abs}_{(1,\omega,\mathbf{k_{\perp}})} &=& i \int d^4 x \,\sqrt{-g} \, j^\mu_R \, A_\mu^{(1,\omega,\mathbf{k_{\perp}})}\\
 &=& \frac{i q}{2\pi^2 k_\perp} \left(\frac{\sinh(\pi \omega/a)}{a}\right)^{1/2} K_{i \omega/a}\left(\frac{k_{\perp}}{a}\right)\nonumber\\
 & & \quad \times \int_{-\infty}^{\infty} d\lambda \; \left[u_x(\lambda\,)k_y-u_y(\lambda)\,k_x\right] \; e^{i \mathbf{k_{\perp}}\cdot\mathbf{x_{\perp}}(\lambda)-i\omega\lambda} \\
{\cal A}^{abs}_{(2,\omega,\mathbf{k_{\perp}})} &=& i \int d^4 x \,\sqrt{-g} \, j^\mu_R \, A_{\mu}^{(2,\omega,\mathbf{k_{\perp}})}\\
 &=& \frac{i q}{2\pi^2} \left(\frac{\sinh(\pi \omega/a)}{a}\right)^{1/2} K'_{i \omega/a}\left(\frac{k_{\perp}}{a}\right) \int_{-\infty}^{\infty} d\lambda \;e^{i \mathbf{k_{\perp}}\cdot\mathbf{x_{\perp}}(\lambda)-i\omega\lambda}
\end{eqnarray}
As in the special case discussed in section \ref{Special case in Rindler frame},  the rate $P^{total}_{R,\mathbf{k_\perp}} $ for Rindler photons with fixed transverse momentum $\mathbf{k_\perp}$ in the Rindler frame with a background thermal bath at Unruh temperature, $T_U=a/2\pi$ is 
\begin{eqnarray}
P^{total}_{R,\mathbf{k_\perp}} &=& \sum_{\ell =1,2} \int_0^\infty d\omega \, \frac{|{\cal A}^{abs}_{(\ell,\omega,\mathbf{k_{\perp}})}|^2}{\Delta \tau} \coth\left(\frac{\pi\omega}{a}\right)
\label{totalrategen}
\end{eqnarray}
Now, we write the total rate corresponding to the polarizations $\ell=1$ and  $\ell=2$ separately as,
\begin{eqnarray}
P^{total}_1 &=& \frac{q^2}{\Delta \tau}\int_{-\infty}^{\infty} d\lambda \int_{-\infty}^{\infty} d\lambda'  \,\left[u_x\, k_y-u_y \,k_x\right] \left[u_x'\,k_y-u_y'\,k_x\right] \frac{e^{i \mathbf{k_{\perp}}\cdot(\mathbf{x_{\perp}}-\mathbf{x'_{\perp}})}}{4\pi^4 k_\perp^2 a} \nonumber\\
 & & \quad \times \int_{0}^{\infty} d\omega \; e^{-i\omega(\lambda-\lambda')} \cosh\left(\frac{\pi\omega}{a}\right) {\left[K_{i \omega/a}\left(\frac{k_{\perp}}{a}\right)\right]}^2\label{total rate for polarization 1}\\
P^{total}_2 &=& \frac{q^2}{\Delta \tau}\int_{-\infty}^{\infty} d\lambda \int_{-\infty}^{\infty} d\lambda'  \, \frac{e^{i \mathbf{k_{\perp}}\cdot(\mathbf{x_{\perp}}-\mathbf{x'_{\perp}})}}{4\pi^4 a}\nonumber\\
 & & \quad \times \int_{0}^{\infty} d\omega \; e^{-i\omega(\lambda-\lambda')} \cosh\left(\frac{\pi\omega}{a}\right) {\left[K'_{i \omega/a}\left(\frac{k_{\perp}}{a}\right)\right]}^2 \label{total rate for polarization 2}
\end{eqnarray}
where, $u_x'$, $u_y'$ and $\mathbf{x'_{\perp}}$ are functions of $\lambda'$ whereas the prime over the Bessel function $K$ denotes the derivative with respect to its argument. Now to obtain the total rate completely in terms of the variables in inertial frames, as in Eq.(\ref{Emission rate for general trajectory}), we first eliminate the Rindler photon frequency $\omega$ from above expressions by performing the $\omega$ integrals. These integrations lead to the Bessel functions of second kind $Y_n(z)$. Then writing these Bessel functions in their integral representations and changing the variable of integration to $\bar{k_z}$, we arrive at the expression of the rate of emission of photons in inertial frame as given in Eq.(\ref{Emission rate for general trajectory}). The detailed calculation is shown below:

The integral over $\omega$ in Eq.(\ref{total rate for polarization 1}) can be simply evaluated to be,
\begin{eqnarray}
\int_{0}^{\infty} d\omega \; e^{-i\omega\sigma} \cosh\left(\frac{\pi\omega}{a}\right) {\left[K_{i \omega/a}\left(\frac{k_{\perp}}{a}\right)\right]}^2 &=& -\frac{\pi^2 a}{4}\, Y_0 \left[\frac{2 k_\perp}{a} \sinh(a\sigma/2)\right]\label{omega integral of Bessel function}
\end{eqnarray}
where, $\sigma=\lambda-\lambda'$ and $Y_n(x)$ is the Bessel function of second kind \cite{Tables of Integrals}.
Now using the relation,
\begin{eqnarray}
{\left[K'_{\nu}\left(z\right)\right]}^2 &=& \frac{1}{2} \left[\frac{d^2}{dz^2}[K_\nu(z)]^2+\frac{1}{z}\frac{d}{dz}[K_\nu(z)]^2-2\left(1+\frac{\nu^2}{z^2}\right)[K_\nu(z)]^2\right]
\end{eqnarray}
the integral over $\omega$ in Eq.(\ref{total rate for polarization 2}) can be obtained to be,
\begin{eqnarray}
\int_{0}^{\infty} d\omega \; e^{-i\omega\sigma} \cosh\left(\frac{\pi\omega}{a}\right) {\left[K'_{i \omega/a}\left(\frac{k_{\perp}}{a}\right)\right]}^2 &=& \frac{\pi^2 a}{8}\, \cosh(a\sigma) Y_0 \left[\frac{2 k_\perp}{a} \sinh(a\sigma/2)\right]\nonumber\\
 & &  +\frac{\pi^2 a}{8}\, Y_2 \left[\frac{2 k_\perp}{a} \sinh(a\sigma/2)\right]\label{omega integral of derivative of Bessel function}
\end{eqnarray}
Substituting Eq.(\ref{omega integral of Bessel function}) and (\ref{omega integral of derivative of Bessel function}) in Eq.(\ref{total rate for polarization 1}) and (\ref{total rate for polarization 2}), the rate $P^{total}_{R,\mathbf{k_\perp}} $ of Rindler photons with fixed transverse momentum $\mathbf{k_\perp}$ becomes,
\begin{eqnarray}
P^{total}_{R,\mathbf{k_\perp}} &=& \frac{q^2}{32 \pi^2 \Delta \tau} \int_{-\infty}^{\infty} d\lambda \int_{-\infty}^{\infty} d\lambda' \,e^{i \mathbf{k_{\perp}}\cdot(\mathbf{x_{\perp}}-\mathbf{x'_{\perp}})}\times\nonumber\\
& & \left[\cosh(a\sigma)\, Y_0(z)+Y_2(z)-\frac{2}{k_\perp^2} \left[u_x k_y-u_y k_x\right] \left[u_x' k_y-u_y' k_x\right] Y_0(z) \right]\label{Total rate with bessel functions}
\end{eqnarray}
where $z=2 k_\perp\sinh(a\sigma/2)/a$. Now we write $Y_0(z)$ as an integral using the integral representation of the Bessel function \cite{Tables of Integrals},
\begin{eqnarray}
Y_\nu(x) &=& -\frac{2}{\pi} \int_0^\infty \cos\left(x \cosh t-\frac{\nu\pi}{2}\right) \cosh(\nu t) dt \qquad \left[|Re(\nu)|<1\right]
\end{eqnarray}
and using the series expansion for the derivative of Bessel function,
\begin{eqnarray}
Y_\nu^{(k)}(x) &=& \frac{1}{2^k} \sum_{n=0}^{k} (-1)^n \binom{k}{n} Y_{\nu-k+2n}(x)
\end{eqnarray}
we express $Y_2(x)$ in terms of $Y_0(x)$ as,
\begin{eqnarray}
Y_2(x) &=& Y_0(x) + 2 \,\dfrac{d^2}{dx^2}Y_0(x).
\end{eqnarray} 
Then identifying $\cosh t$ as $\bar{k_0}/k_\perp$, where $\bar{k_0}=\sqrt{\bar{k_z}^2+k_\perp^2}$, with $\bar{k_0}$ given by Eq.(\ref{transformed frequency}) and the corresponding $\bar{k_z}$ given by Eq.(\ref{transformed momentum}), we get the Bessel functions $Y_0(z)$ and $Y_2(z)$ as the following integrals with $\bar{k_z}$ as integration variable.
\begin{eqnarray}
Y_0 \left[\frac{2 k_\perp}{a} \sinh(a\sigma/2)\right] &=& \frac{-1}{\pi} \int_{-\infty}^{\infty} \cos\left(\frac{2\bar{k_0}}{a} \sinh(a\sigma/2)\right) \frac{d\bar{k_z}}{\bar{k_0}} \\
Y_2 \left[\frac{2 k_\perp}{a} \sinh(a\sigma/2)\right] &=& \frac{1}{\pi} \int_{-\infty}^{\infty} \cos\left(\frac{2\bar{k_0}}{a} \sinh(a\sigma/2)\right) \left(\frac{\bar{k_z}^2+\bar{k_0}^2}{k_\perp^2}\right) \frac{d\bar{k_z}}{\bar{k_0}}
\end{eqnarray}
Substituting these integral representations of $Y_0(z)$ and $Y_2(z)$ in Eq.(\ref{Total rate with bessel functions}), we get for the rate $P^{total}_{R,\mathbf{k_\perp}}$,
\begin{eqnarray}
P^{total}_{R,\mathbf{k_\perp}} &=& \frac{q^2}{32 \pi^3 k_\perp^2 \Delta \tau} \int_{-\infty}^{\infty} d\lambda \int_{-\infty}^{\infty} d\lambda' \,e^{i \mathbf{k_{\perp}}\cdot(\mathbf{x_{\perp}}-\mathbf{x'_{\perp}})} \int_{-\infty}^{\infty} \frac{d\bar{k_z}}{\bar{k_0}} \cos\left(\frac{2\bar{k_0}}{a} \sinh(a\sigma/2)\right) \nonumber\\
 & & \times \left[-k_\perp^2 \cosh(a\sigma)+\bar{k_z}^2+\bar{k_0}^2+2 \left[u_x k_y-u_y k_x\right] \left[u_x' k_y-u_y' k_x\right]\right]
\end{eqnarray}
Now, as the current $j^\mu$ is conserved, i.e. $\partial_\mu j^\mu=0$, one can write for its Fourier transform,
\begin{eqnarray}
\int d^4 x \sqrt{-g}\, e^{i k_\nu x^\nu} \,j^\mu(x) \,k_\mu &=& 0
\end{eqnarray}
In Minkowski frame for the current given by Eq.(\ref{current in Minkowski for general trajectory}), this equation simplifies to
\begin{eqnarray}
\int_{-\infty}^{\infty} d\lambda \, e^{i \mathbf{k_{\perp}}\cdot(\mathbf{x_{\perp}}-\mathbf{x'_{\perp}})} \exp\left(\frac{2i\bar{k_0}}{a} \sinh(a\sigma/2)\right) u^\mu(\lambda) k_\mu &=& 0\label{Constraint on four velocity}
\end{eqnarray}
The above constraint on $u^\mu(\lambda)$ with the definitions of $\bar{k_0}$ and $\bar{k_z}$ simplifies the expression of total rate with the term in the bracket reducing to $-2 k_\perp^2 u^\mu(\lambda) u_\mu(\lambda')$. The total rate is then given by,
\begin{eqnarray}
P^{total}_{R,\mathbf{k_\perp}} &=& \frac{q^2}{\Delta \tau} \int_{-\infty}^{\infty}\frac{d\bar{k_z}}{(2\pi)^3 \, 2\bar{k_0}} \int_{-\infty}^{\infty} d\lambda \int_{-\infty}^{\infty} d\lambda' \, \left[-u^\mu(\lambda) u_\mu(\lambda')\right] \nonumber\\
& & \times \,e^{i \mathbf{k_{\perp}}\cdot(\mathbf{x_{\perp}}-\mathbf{x'_{\perp}})} \cos\left(\frac{2\bar{k_0}}{a} \sinh(a\sigma/2)\right)
\end{eqnarray}
The overall rate is then given by integrating over the transverse momenta $k_x$ and $k_y$. Using the symmetry in $\lambda$ and $\lambda'$ in the above expression and interchanging the limits of the $k_x$, $k_y$ integrals, it can be expressed as 
\begin{eqnarray}
P^{total}_{R}  &=& \int_{-\infty}^{\infty} dk_x \int_{-\infty}^{\infty} d k_y \,P^{total}_{R,\mathbf{k_\perp}} \nonumber \\
&=& \frac{-q^2}{\Delta \tau}\, \int_{-\infty}^{\infty} dk_x \int_{-\infty}^{\infty} d k_y \int_{-\infty}^{\infty}\frac{d\bar{k_z}}{(2\pi)^3 \, 2\bar{k_0}} \int_{-\infty}^{\infty} d\lambda \int_{-\infty}^{\infty} d\lambda' \, \left[u^\mu(\lambda) u_\mu(\lambda')\right] \nonumber\\
& & \times \,\cos\left(\mathbf{k_{\perp}}\cdot(\mathbf{x_{\perp}}-\mathbf{x'_{\perp}})\right) \exp\left(\frac{2i\bar{k_0}}{a} \sinh(a\sigma/2)\right) \nonumber\\
&=& \frac{-q^2}{\Delta \tau}\, \int_{-\infty}^{\infty} dk_x \int_{-\infty}^{\infty} d k_y \int_{-\infty}^{\infty}\frac{d\bar{k_z}}{(2\pi)^3 \, 2\bar{k_0}} \int_{-\infty}^{\infty} d\lambda \int_{-\infty}^{\infty} d\lambda' \, \left[u^\mu(\lambda) u_\mu(\lambda')\right] \nonumber\\
& & \times \,e^{i \mathbf{k_{\perp}}\cdot(\mathbf{x_{\perp}}-\mathbf{x'_{\perp}})} \exp\left(\frac{2i\bar{k_0}}{a} \sinh(a\sigma/2)\right)\label{total rate for an arbitrary trajectory}
\end{eqnarray}
which matches with the expression of emission rate obtained in inertial frame in Eq.(\ref{Emission rate for general trajectory}). We proceed to calculate the number of emitted photons with the classical framework in the next section.

\subsection{Radiation using classical electrodynamics}\label{Classical calculation for general case}
In this subsection, we perform a straightforward classical electrodynamics calculation involving the spectral angular distribution of radiation from an accelerating charge on the Rindler trajectory with arbitrary transverse motion as described by Eq.(\ref{velocity in Rindler frame for general trajectory}).

The expression for number of emitted Minkowski photons presented in Eq.(\ref{Total number of emitted photons}) is, in general, valid for any arbitrary trajectory of charge $q$. The vector $\bar{F}(k_x,k_y,k_z) $ defined in Eq.(\ref{Vector F}), then can be re-expressed as,
\begin{eqnarray}
F^{l}(k_x,k_y,k_z)  &=& \epsilon^{lmn} \,\hat{r}_m \,\int_{-\infty}^{\infty} d\lambda {\left(\frac{dr_q}{d\lambda}\right)}_n \exp\left(\,-ik_0\,[\,\hat{r}\cdot\bar{r_q}(\lambda)-t(\lambda)\,]\,\right)
\end{eqnarray}
where, $\epsilon^{lmn}$ is the completely anti-symmetric Levi-Civita symbol in three dimensions. Then for the magnitude of vector $\bar{F}(k_x,k_y,k_z) $, we get
\begin{eqnarray}
{\left|\bar{F}(k_x,k_y,k_z)  \right|}^2 &=& \delta_{al} F^{a} {F^l}^{*} \nonumber \\
 &=& \delta_{al}\, \epsilon^{abc}\, \epsilon^{lmn} \,\hat{r}_b\,\hat{r}_m \int_{-\infty}^{\infty} d\lambda \int_{-\infty}^{\infty} d\lambda' {\left(\frac{d\bar{r_q}(\lambda)}{d\lambda}\right)}_c {\left(\frac{d\bar{r_q}(\lambda')}{d\lambda'}\right)}_n \nonumber\\
 & & \times \exp\left(\,-ik_0\,\left[\,\hat{r}\cdot\left(\bar{r_q}(\lambda)-\bar{r_q}(\lambda')\right)-\left(t(\lambda)-t(\lambda')\right)\,\right]\,\right) \nonumber \\
 &=& \int_{-\infty}^{\infty} d\lambda \int_{-\infty}^{\infty} d\lambda' \exp\left(\,-ik_0\,\left[\,\hat{r}\cdot\left(\bar{r_q}(\lambda)-\bar{r_q}(\lambda')\right)-\left(t(\lambda)-t(\lambda')\right)\,\right]\,\right)\nonumber \\
 & &\times \left[\left(\hat{r}\cdot\hat{r}\right)\left(\frac{d\bar{r_q}(\lambda)}{d\lambda}\cdot\frac{d\bar{r_q}(\lambda')}{d\lambda'}\right)-\left(\hat{r}\cdot\frac{d\bar{r_q}(\lambda)}{d\lambda}\right)\left(\hat{r}\cdot\frac{d\bar{r_q}(\lambda')}{d\lambda'}\right)\right]\label{magnitude of vector F}
\end{eqnarray}
Here we have used the identity, $\delta_{al}\, \epsilon^{abc}\, \epsilon^{lmn} = \left(\delta^{bm}\delta^{cn}-\delta^{bn}\delta^{cm}\right)$ to arrive at the final equality. Now, using the constraint on $u^\mu$ as defined in Eq.(\ref{Constraint on four velocity}) and the definition of unit direction vector $\hat{r}=\bar{k}/k_0$, the term $(d\bar{r_q}/d\lambda)\cdot\hat{r}$ can be replaced by $(dt/d\lambda)$. The magnitude $|\bar{F}(k_x,k_y,k_z) |^2$ then becomes,
\begin{eqnarray}
{\left|\bar{F}(k_x,k_y,k_z)  \right|}^2 &=& \int_{-\infty}^{\infty} d\lambda \int_{-\infty}^{\infty} d\lambda' \exp\left(\,-i\,\left[\,\bar{k}\cdot\left(\bar{r_q}(\lambda)-\bar{r_q}(\lambda')\right)-k_0\left(t(\lambda)-t(\lambda')\right)\,\right]\,\right)\nonumber \\
 & &\quad \times \; \left[\frac{d\bar{r_q}(\lambda)}{d\lambda}\cdot\frac{d\bar{r_q}(\lambda')}{d\lambda'}-\frac{dt(\lambda)}{d\lambda}\frac{dt(\lambda')}{d\lambda'}\right]
\end{eqnarray}
Now, substituting the trajectory, $\bar{r_q}(\lambda)= x(\lambda) \,\hat{i}+y(\lambda) \,\hat{j}+a^{-1} \cosh(a\lambda)\, \hat{k}$, with the time co-ordinate as $t(\lambda)=a^{-1} \sinh(a\lambda)$, we get,
\begin{eqnarray}
{\left|\bar{F}(k_x,k_y,k_z)  \right|}^2 &=& \int_{-\infty}^{\infty} d\lambda \int_{-\infty}^{\infty} d\lambda' \,\left[\,-u^\mu(\lambda)u_\mu(\lambda')\,\right]\,e^{-i \mathbf{k_{\perp}}\cdot(\mathbf{x_{\perp}}-\mathbf{x'_{\perp}})} \nonumber\\
 & & \times \,\exp\left[\frac{2i\bar{k_0}}{a} \sinh\left(\frac{a\sigma}{2}\right)\right]  
\end{eqnarray}
where $\bar{k_0}$ is as defined in Eq.(\ref{transformed frequency}). With this expression for ${\left|\bar{F}(k_x,k_y,k_z)  \right|}^2$, the total number of emitted photons $N_{M} $ is obtained to be,
\begin{eqnarray}
N_{M} &=& \int_{-\infty}^{\infty} dk_x \int_{-\infty}^{\infty} dk_y \int_{-\infty}^{\infty} \frac{dk_z}{k_0}\; \frac{q^2}{4\pi^2}\, {\left|\bar{F}(k_x,k_y,k_z)  \right|}^2 \nonumber \\
 &=&\frac{-q^2}{4\pi^2} \int_{-\infty}^{\infty} \frac{dk_z}{k_0} \int_{-\infty}^{\infty} d\lambda \int_{-\infty}^{\infty} d\lambda' \,\left[\,u^\mu(\lambda)u_\mu(\lambda')\,\right]\nonumber\\
 & & \times \,e^{-i \mathbf{k_{\perp}}\cdot(\mathbf{x_{\perp}}-\mathbf{x'_{\perp}})}\,\exp\left(\frac{2i\bar{k_0}}{a} \sinh\left(\frac{a\sigma}{2}\right)\right) \nonumber \\
  &=& 4\pi (-q^2) \int_{-\infty}^{\infty} dk_x \int_{-\infty}^{\infty} dk_y \int_{-\infty}^{\infty} \frac{dk_z}{(2\pi)^3 2k_0} \int_{-\infty}^{\infty} d\lambda \int_{-\infty}^{\infty} d\lambda' \,\left[\,u^\mu(\lambda)u_\mu(\lambda')\,\right]\nonumber\\
 & & \times \,e^{-i \mathbf{k_{\perp}}\cdot(\mathbf{x_{\perp}}-\mathbf{x'_{\perp}})}\,\exp\left(\frac{2i\bar{k_0}}{a} \sinh\left(\frac{a\sigma}{2}\right)\right)
\end{eqnarray}
With a comparison of the above expression with Eq.(\ref{total rate for an arbitrary trajectory}), one can simply write the number of emitted photons in terms of total rate of a Rindler detector as,
\begin{eqnarray}
N_{M} &=& 4\pi\, \Delta\tau\, P^{total}_{R}
\end{eqnarray}
As discussed in the previous section \ref{Classical calculation of special case}, the proportionality constant is $4\pi\, \Delta\tau$, which will be finite for a charge accelerating for a finite time.

\section{Discussion} \label{discussion}
An equivalence is demonstrated, by an explicit quantum calculation till linear order in perturbations, between the Minkowski photon emission rate, in the inertial frame, for an accelerating charge moving on a Rindler trajectory with additional arbitrary transverse motion and the combined Rindler photon emission and absorption rate of the same charge in the Rindler frame in the presence of the Davies Unruh bath. The equivalence also extends, between the Bremsstrahlung emitted by the same charge as calculated using the machinery of classical electrodynamics and the expectation values from the quantum calculations mentioned above.

Two immediate observations follow: (i) as noted in \cite{Classical ED}, the presence of the Unruh bath in the Rindler frame is utmost necessary for the quantum calculation to match with the classical one or with the quantum calculation in the inertial frame.
The particular $\cosh{(\omega/2T)}$ factor in Eqs.(\ref{totalrate}) and (\ref{totalrategen}) which arises through the combined emission and absorption rate is crucial for the equivalence to work. The $\cosh{(\omega/2T)}$ factor is universal in the sense that it is common for all trajectories under consideration and does not require any fudging. In retrospect, it re-affirms earlier observations that absorption of Rindler particles in the Rindler frame appears to the inertial observer as emission of Minkowski particles in the inertial frame. Additionally, one needs to clarify that even though the Unruh effect is purely quantum in its origin, the said equivalence works due to the cancellation of the $\hbar$ in the thermal Planckian factor (leading to the $\cosh{(\omega/2T)}$ factor) since both the photon energies and Unruh temperature are linearly proportional to $\hbar$. Any other value of the thermal bath temperature chosen shall not suffice. In fact, once the choice is made, the form of classical mode solutions in the Minkowski and Rindler frames are sufficient (mathematically) to arrive at the results. This strongly suggests that the Unruh effect is necessary to explain certain classical features of electromagnetic radiation when two different frames are involved. (ii) The correspondence principle in quantum mechanics holds for many particle systems in the limit when the number of particles involved are large. In the present scenario, the charge is taken to be classical and moving on a well defined trajectory. It is however coupled to the quantised electromagnetic field. In linear order perturbation theory, only a single photon excitation or de-excitation is admissible, although it holds true for every (continuum of) energy eigenstates of the quantised Maxwell fields. In principle, the point charge does excite an infinite number of photons of the whole frequency spectrum due to its motion and hence one could argue that the large particle limit is inherently built into the case under investigation. Then, it should be no surprise that quantum calculation in inertial frame or the Rindler frame agrees with that of Larmor radiation using classical electrodynamics. However, it would be interesting to go further than the first order approximation to analyse the quantum effects of the Davies - Unruh effect beyond the classicality of Maxwell's equations. It would be interesting to test the equivalence for a more broad class of trajectories wherein additional motion along the special direction $\xi$ is also allowed. 

In terms of experimental prospects, our results have generalised the proposal of Cozzella et al. \cite{Classical ED} for the virtual confirmation of the Unruh effect using classical Bremsstrahlung. The requirement of circular motion in the transverse directions could be relaxed depending on experimental set-up constraints etc or any slight deviation from the circularity will not affect the main conclusions of the proposal since the equivalence is shown to hold for arbitrary transverse motion.  

\section*{Acknowledgments}

SK and KP thank the Department of Science and Technology, India, for financial support.

\end{document}